\documentclass[doublecol]{epl2}

\usepackage{graphicx}
\usepackage{dcolumn}
\usepackage{bm}

\title{Interaction induced trapping and pulsed emission of a magnetically insensitive Bose-Einstein Condensate}
\shorttitle{Trapping and emission of a magnetically insensitive BEC} 
\author{S. Middelkamp\inst{1} \and I. Lesanovsky\inst{2} \and P. Schmelcher\inst{1,3}}
\shortauthor{S. Middelkamp \etal}

\institute{    
\inst{1} Theoretische Chemie, Physikalisch-Chemisches Institut, Universit\"at Heidelberg, INF 229, 69120 Heidelberg, Germany\\
\inst{2} Institut f\"ur Theoretische Physik, Universit\"at Innsbruck, Austria\\
\inst{3} Physikalisches Institut, Universit\"at Heidelberg, Philosophenweg 12, 69120 Heidelberg, Germany\\
}
\date{\today}


\abstract{
We demonstrate that atoms in magnetically insensitive hyperfine states (m=0) can be trapped efficiently by a Bose-Einstein Condensate of the same atomic species occupying a different hyperfine state. The latter is trapped magnetically. Hyperfine state changing collisions, and therefore loss of the trapped (m=0) atoms, are shown to be strongly inhibited in case of a low density of the confined atomic cloud. We monitor the transition from a 'soft' to a 'hard' effective potential by studying the backaction of the trapped (m=0) atoms onto the condensate which provides their confinement. The controlled outcoupling of the trapped atoms by shaping the condensate's wavefunction is explored. We observe a pulsed emission of atoms from the trapping region reminiscent of an atom laser.
}
\begin{document}
\maketitle

Trapped ultracold atomic gases offer outstanding possibilities to model complex quantum systems. A paradigm is the Mott-Insulator phase transition demonstrated with atoms trapped in a lattice potential \cite{Greiner02}. Thus, a key ingredient for advancing the possibilities to explore the quantum dynamics of many body ensembles with ultracold atoms is to improve existing methods or to find novel ways for the control of the external motion of atoms. Usually static or time-dependent electro-magnetic fields are employed for this task \cite{Folman,Grimm,Wiemann,Fortagh,Dalfovo}. A prominent example are optical lattices formed by counter propagating light waves who facilitated the observation of effects like the above-mentioned Mott-Insulator phase transition \cite{Greiner02}, second-order tunneling \cite{Foelling07} or Josephson oscillations \cite{Albietz05}. In a similar fashion so-called radio-frequency dressed adiabatic potentials that emerge from a combination of static and oscillating magnetic fields \cite{Lesanovsky06,Hofferberth06} have been used in order to coherently manipulate matterwaves \cite{Schumm05} and to investigate the decoherence dynamics of one-dimensional Bose gases \cite{Hofferberth07}.

Magnetic trapping relies on the coupling of the total angular momentum $\mathbf{F}$ to the magnetic field vector. The resulting potential is proportional to the magnetic projection quantum number $m$ ranging from $-F$ to $F$. Hence, in the linear Zeeman regime the $m=0$ state is insensitive to the magnetic field and can therefore not be trapped magnetically. By contrast, in optical traps it is possible to trap the $m=0$ component and to achieve atomic spinor Bose-Einstein Condensates (BEC) where the different spin components can exchange population coherently \cite{Mur-Petit06}. This spin mixing dynamics has been studied theoretically \cite{Ho98,Ohmi98} and experimentally, e.g. for $^{23}$Na \cite{Stenger} and $^{87}$Rb \cite{Schmaljohann}.

In this letter we show that the $m=0$ component can nevertheless be confined within a magnetic trap. However, the confinement is not provided by a potential due to the magnetic field but rather by a BEC formed by atoms in a state with $m\neq 0$. We show that for a large BEC density in the $m\neq0$ state the atoms in the $m=0$ state behave as if they were confined by a potential which is constituted by the condensate density. Eventually we demonstrate how atoms in the $m=0$ state are released in a controlled manner from the 'trap' if the BEC density is varied. This method complements current methods to outcouple atoms from a BEC, e.g. radio frequency \cite{Mewes97} or Raman transitions \cite{Hagley99} and facilitates the creation of an atom laser for $m=0$ atoms.

In the following we consider atoms in a $F=1$ hyperfine manifold and restrict ourselves to an effective one-dimensional description accounting only for the longitudinal dynamics. The transversal confinement may be provided by an isotropic harmonic potential with a sufficiently high trap frequency $\omega_\perp$ (associated with the harmonic oscillator length $a_\perp$) such that the transversal dynamics is frozen out (see chap. 1 in \cite{Kevrekidis} and ref. \cite{perez-garcia, Jackson}) i.e. only the corresponding ground state is occupied. The longitudinal motion takes place in $x$-direction and the corresponding mean field equations for the evolution of the spinor components $\Psi_m=\Psi_m(x,t)$ are given by
\begin{eqnarray}
  i\hbar \partial_t \Psi_0&=&\left[-\frac{\hbar^2}{2M} \partial^2_x+V_0(x)+g_0 n\right]\Psi_0\nonumber\\
&&+\frac{g_1}{\sqrt{2}}\left[G_+\Psi_{+1}+G_-\Psi_{-1}\right]\nonumber\\
i\hbar \partial_t \Psi_{\pm 1}&=&\left[-\frac{\hbar^2}{2M} \partial^2_x+V_{\pm 1}(x)+g_0 n\right]\Psi_{\pm 1}\nonumber\\
&&+g_1\left[\frac{1}{\sqrt{2}}G_{\mp}\Psi_0\pm G_z\Psi_{\pm 1}\right]\label{eqn:spinor_equations}
\end{eqnarray}
with the atomic mass $M$, the total atomic density $n=\sum_{m=-1}^1 \left|\Psi_m\right|^2$, $G_z=\left|\Psi_{+1}\right|^2-\left|\Psi_{-1}\right|^2$ and $G_+=G_-^*=\sqrt{2}\left[\Psi_1^*\Psi_0+\Psi_0^*\Psi_{-1}\right]$. The coupling constants are $g_0=2\hbar^2(a_0+2a_2)/(3Ma^2_\perp)$ and $g_1=2\hbar^2(a_2-a_0)/(3Ma^2_\perp)$ where $a_{0/2}$ are the s-wave scattering lengths for the scattering-channels with total spin $0$ and $2$. The particle number in the respective hyperfine state is calculated according to $N_m=\int\,dx\left|\Psi_m\right|^2$. In this work we consider $^{23}$Na with the respective scattering lengths $a_0=2.43\, \mathrm{nm}$ and $a_2=2.75\, \mathrm{nm}$. The axial trap frequency is chosen $\omega_\perp=2$ kHz which gives rise to a transverse oscillator length $a_\perp=2.97\times 10^{3}\, \mathrm{nm}$.

The spin-dependent potential under consideration is of the form $V_m(x)=m V(x)$ with $V(x)$ forming a double-well. As shown in Refs. \cite{Lesanovsky06,Lesanovsky06b} such a potential can be created using a standard magnetic trap of Ioffe-Pritchard type which is dressed by a homogeneous radio-frequency field. Instead of using the actual form of $V(x)$ which is provided in Ref. \cite{Lesanovsky06} we rather model the double-well as $V(x)=\gamma+\eta \exp\left(-x^2/\sigma^2\right)+(1/2) M\omega^2 x^2$. This model potential, which is easily implemented numerically, covers the important features of the exact potential and can be made to match the experimental situation by tuning the parameters $\gamma$, $\eta$, $\sigma$ and $\omega$. For our calculations we choose $\gamma=0$, $\eta=1.05\times 10^{30}\, \mathrm{J} $, $\sigma=28\, \mu\mathrm{m} $ and $\omega=71\, \mathrm{Hz}$.
\begin{figure}
\includegraphics[angle=0,width=5cm]{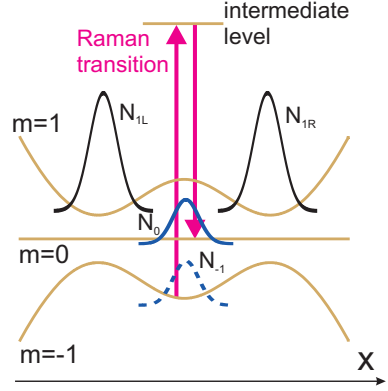}
\caption{Preparation of the initial state: BECs prepared in the two wells of the $m=1$ manifold and the single well of the $m=-1$ potential. All $m=-1$ atoms are transferred by a Raman transition to the $m=0$ state.}\label{fig:setup}
\end{figure}
A sketch of this potential is shown in fig. \ref{fig:setup}. While atoms in the $m=1$ state are subjected to an overall confining double well potential, atoms in the state $m=-1$ can be trapped only temporarily in metastable states of a single well which is unbounded from below. We assume this state to be stable over the timescale of interest. Additionally, the atom numbers and well separations are chosen such that there is negligible initial overlap between the three atomic clouds located in the three wells. The initial state for our investigations is created by transferring the entire population of the $m=-1$ into the $m=0$ state which experiences no confining potential. Such a transfer can be achieved experimentally by two sufficiently broad banded microwave or laser pulses providing a Raman transition via intermediate excited states.

We prepare our initial state by relaxation, i.e. imaginary time propagation of a trial wave function. Since there is no initial overlap between the wavepackets each of them can be prepared independently. We therefore introduce the number of particles confined in the left and right well by $N_{1L}=\int_{-\infty}^0 \left|\Psi_1\right|^2$ and $N_{1R}=\int^{\infty}_0 \left|\Psi_1\right|^2$, respectively (see fig. \ref{fig:setup}). They obey $N_{1L}+N_{1R}=N_1$. After having obtained the ground states in each potential well we set $\Psi_0=\Psi_{-1}$ and subsequently put $\Psi_{-1}=0$ assuming a perfect population transfer. Then the system is evolved in time according to eqs. (\ref{eqn:spinor_equations}) using the Adams-Bashforth-Moulton predictor-corrector method.

Throughout this work we will consider the regime of low densities of the $m=0$ atoms and a small maximum overlap region of the $m=0$ and $m=1$ wavepackets in which case $\partial_t N_{-1}$ becomes negligibly small. For analytical considerations we therefore make the very good approximation $\Psi_{-1}=0$ for all times (see below for a corresponding discussion) and arrive at the two equations of motion: 
\begin{eqnarray}
  i\hbar\partial_t \Psi_0&=&\left[-\frac{\hbar^2}{2M}\partial_x^2+g_0\left|\Psi_0\right|^2+(g_0+g_1)\left|\Psi_{1}\right|^2\right]\Psi_0\label{eq:psi0}\\
i\hbar\partial_t \Psi_1&=&\left[-\frac{\hbar^2}{2M}\partial_x^2+V(x)+(g_0+g_1)(\left|\Psi_0\right|^2+\left|\Psi_1\right|^2)\right]\Psi_1\label{eq:psi1}
\end{eqnarray}
i.e. the $m=-1$ component is not involved in the time-evolution at all.
\begin{figure}[htb]
\begin{center}
\includegraphics[angle=270,width=7cm]{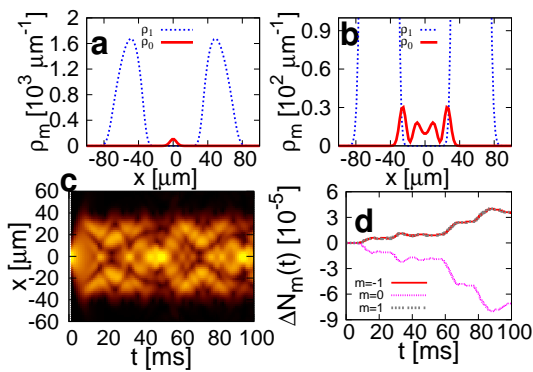}
     \end{center}
\caption{\textbf{a}: Density of the individual $m$ components ($m=1$ dashed line, $m=0$ solid line) of the initial state with $N_1=10^{5}$ ($N_{1R}=N_{1L}$) and $N_0=10^3$. \textbf{b}: Density snapshot after the time $t=13$ ms. \textbf{c}: Time evolution of the density of the $m=0$ component. \textbf{d}: Relative change of the number of particles  $\triangle N_{m}(t)=\frac{N_m(t)-N_m(0)}{N_{0}(0)}$ for $N_0(0)=10^3$. The change of the particle number of the $m=1$ and $m=-1$ components are equal (upper lines). The lower line denotes the change of the particle number of the $m=0$ component.}\label{fig:trapping}
\end{figure}
From eq. (\ref{eq:psi0}) we observe that, although there is no external trap acting on $\Psi_{0}$, we can identify the term $V_{eff}(x)=(g_0+g_1)\left|\Psi_{1}\right|^2$ as an effective potential. However, this potential depends explicitly on the $m=1$ density and implicitly on the $m=0$ density as seen from eq. (\ref{eq:psi1}). In order to demonstrate that $V_{eff}(x)$ can provide trapping we prepare an initial state with $N_{1R}=N_{1L}=(1/2)\times 10^5$ and $N_0=10^3$ as shown in fig. \ref{fig:trapping}a. During the first several milliseconds the non-stationary $m=0$ wave packet broadens until parts of it hit the atoms in the $m=1$ state and $\Psi_1$ and $\Psi_0$ overlap. The situation at this instant of time is depicted in fig. \ref{fig:trapping}b. The $m=0$ atoms are reflected back completely and are thus eventually trapped as displayed in fig. \ref{fig:trapping}c, where the $m=0$ density over a time interval of $100\,\mathrm{ms}$ is shown. As can be seen the atoms are effectively confined to the interval $-50\,\mu \mathrm{m} < x < 50\,\mu \mathrm{m}$. Eq. (\ref{eq:psi0}) shows that the trapping is rooted in the density-density interaction between atoms in different spin states. A similar behavior is therefore expected for $^{87}$Rb and for any other atomic species whose scattering lengths obey $g_0+g_1>0$. Moreover, the density-density interaction implies  that the demonstrated trapping effect should also be observable if two different atomic species are employed instead of two different spin states of the same atom. This, of course, requires that the atomic density of one of the species is shaped accordingly.

Let us now return to the question of particle loss due to the scattering process $(m,m^\prime)=(0,0)\rightarrow(1,-1)$. In fig. \ref{fig:trapping}d we show the temporal evolution of the relative change of the particle number $\triangle N_m(t)=\frac{N_m(t)-N_m(0)}{N_{0}(0)}$ in each of the three $m$-channels for $N_{0}(0)=10^3$. The relative particle loss from the trapped atoms in the $m=0$ state is negligibly small for the time interval $100\,\mathrm{ms}$ and therefore eqs. (\ref{eq:psi0},\ref{eq:psi1}) provide an accurate approximation. The scattering process $(0,0)\rightarrow(1,-1)$, and thus the loss of $m=0$ atoms, is strongly suppressed. For vanishing $m=1$ and $m=-1$ components eqs. (\ref{eqn:spinor_equations}) reduce to the Gross-Pitaevskii equation for the $m=0$ component only and spin exchange processes are therefore absent. The microscopic spin exchange process can therefore in the mean field picture only occur if there is an overlap of at least two different spin components. In our situation this overlap is negligibly small. In order to see this more quantitatively we derive an upper bound for the population rate of the $m=-1$ component on the basis of eq. (\ref{eqn:spinor_equations}). The particle number $N_{-1}$ obeys $\partial_t N_{-1}=\int dx \left[\partial_t \Psi_{-1}\Psi_{-1}^*+\partial_t \Psi_{-1}^*\Psi_{-1}\right]$. For short times the evolution of $\Psi_{-1}$ is governed by $\partial_t \Psi_{-1}=-i\frac{g_1}{\hbar}\Psi_0^2\Psi^*_1$. Assuming the r.h.s. to be constant leads then to $\Psi_{-1}(t)\approx \partial_t \Psi_{-1}\,t$. Considering exclusively the process that populates $\Psi_{-1}$  we find $\partial_t N_{-1}\approx\int dx \partial_t \Psi_{-1}\partial_t \Psi_{-1}^* t\approx (g_1/\hbar)^2\int dx \left|\Psi_0\right|^4\left|\Psi_1\right|^2\,t$. Introducing a typical overlap region  $\triangle x$ between the $m=0$ and $m=1$ wave functions and the corresponding maximal values for the densities $n_{0,\mathrm{max}}$,  $n_{1,\mathrm{max}}$ inside this region, we find the following upper bound for the particle number increase in the $m=-1$ mode:
\begin{eqnarray}
  \partial_t N_{-1}<(\frac{g_1}{\hbar})^2 \triangle x\, n_{0,\mathrm{max}}^2n_{1,\mathrm{max}}t. \label{eq:particle_loss}
\end{eqnarray}
Estimating the increase of the particle number in the $m=-1$ according to eq. (\ref{eq:particle_loss}) we find with $n_{0,max}=25\,\mu m^{-1}$, $n_{1,max}=100\,\mu m^{-1}$ and an overlap of $\triangle x \approx 20\,\mu m$ a rate of $\partial_t N_{-1}<0.02t\,/ (ms)^2$ which is consistent with our numerical data. Beyond this estimate fig. \ref{fig:trapping}d shows a step-like increase of $N_{-1}$ which is due to the time-dependence of the overlap of $\Psi_0$ and $\Psi_1$.

\begin{figure}
\includegraphics[angle=0,width=7cm]{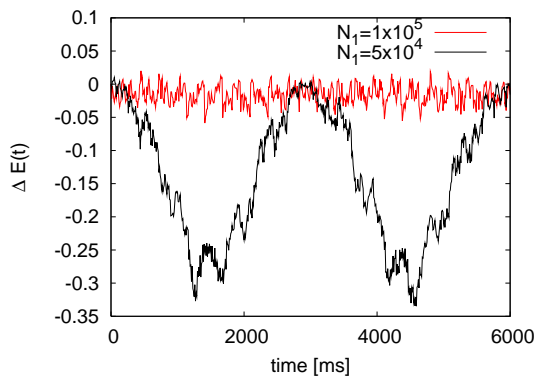}
\caption{Relative energy change $\triangle E(t)=\frac{E_{eff}(t)-E_{eff}(0)}{E_{eff}(0)}$ of the $m=0$ component for $N_0=10^3$, $N_1=10^5$ and $5\times 10^4$ and $N_{1R}=N_{1L}$.}\label{fig:energy}
\end{figure}
Let us now address the question of a backaction of the $m=0$ atoms onto the $m=1$ atoms. As demonstrated above the effective potential $V_{eff}(x)=(g_0+g_1)\left|\Psi_{1}\right|^2$ can grant efficient confinement of the $m=0$ atoms. However, according to the coupled eqs. (\ref{eq:psi0},\ref{eq:psi1})  $\Psi_1$ does not evolve independently of $\Psi_0$. Hence it is expected that the trapped atoms act back on the atoms in the $m=1$ state and thereby modify the trapping potential itself. In order to quantify this effect we calculate the effective energy
$  E_{eff}=\int\,dx \Bigl(-\frac{\hbar^2}{2M}\left|\partial^2_x\Psi_0\right|^2+
\frac{(a_0+2a_2)\hbar^2}{3Ma^2_\perp}\left|\Psi_0\right|^4+\frac{2a_2\hbar^2}{Ma_\perp^2}\left|\Psi_0\right|^2\left|\Psi_1\right|^2\Bigr)$
which is the energy contained in the $m=0$ component itself plus a contribution arising due to the interaction of the $m=0$ with the $m=1$ density. If $\left|\Psi_{1}\right|^2$ was static, i.e. it did not change its shape as a function of time, $E_{eff}=E_{eff}(t)$ would be conserved. Conversely, energy exchange between the spin components will be reflected in a variation of $E_{eff}$ over time. In fig. \ref{fig:energy} we illustrate the time-dependence of the quantity $\triangle E(t)=\frac{E_{eff}(t)-E_{eff}(0)}{E_{eff}(0)}$ for $N_0=10^3$ and two different particle numbers $N_1=10^5$ and $5\times 10^4$. The shape of the initial state is the same as discussed before, i.e. symmetric occupation of the $m=1$ double-well potential (see fig. \ref{fig:trapping}a). We observe that for $N_1=10^5$ the relative energy change is of the order of $5\,\%$ over the shown time interval of $10^4\, ms$. Here the effective potential can be approximately considered as static. The situation changes if $N_1$ is lowered. More concrete, for $N_1=5\times 10^4$ we observe large oscillations of $\triangle E$ with peak values up to $35\,\%$ and the effective potential picture breaks down.

So far we have been focusing on an initial state with a symmetric occupation of the $m=1$ double well. We will now investigate the situation $N_{1L}\neq N_{1R}$ where $N_{1R}$ is not large enough to provide a complete confinement of the $m=0$ atoms.
\begin{figure}[htb]
\begin{center}
\includegraphics[angle=270,width=7cm]{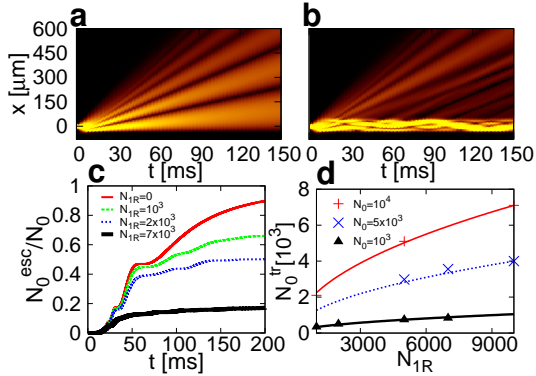}
     \end{center}
\caption{\textbf{a},\textbf{b}: The time evolution of the $m=0$ density for $150$ ms and an initial state with $N_0=10^3$, $N_{1L}=10^5$, $N_{1R}=0$ (\textbf{a}) and $N_{1R}=7\times 10^3$ (\textbf{b}). Atoms are released from the trap region and escape to the $x>0$ halfspace. \textbf{c}: Time evolution of the relative number of escaped $m=0$ atoms for (from top to bottom) $N_{1R}=\lbrace 0 $, $10^3$, $2\times10^3$, $7\times10^3\rbrace$. \textbf{d}: Number of trapped atoms $N_{0}^{tr} = N_{0}-N_0^{esc}$ after $t=150\, ms$ as function of $N_{1R}$ for  $N_0=10^4$ ($+$), $N_0=5\times10^4$ ($\times$) and  $N_0=10^3$ ($\triangle$). The lines were generated by fitting $N_{0}^{tr}=\kappa \sqrt{N_{1R}}$.}
\label{fig:release}
\end{figure}
In fig. \ref{fig:release}a the evolution of the $m=0$ density is shown for the particle numbers $N_0=10^3$, $N_{1L}=10^5$ and $N_{1R}=0$. In the case, where there is no occupation of the right hand well, the atoms are free to leave the trapping region to the $x>0$ halfspace. Escape to the opposite direction is prevented by the $m=1$ wave packet in the left well ($N_{1L}$) and thus atoms which are initially going to the left are reflected. They interfere with the atoms initially going to the right thereby giving rise to an emission of a sequence of distinct wave packets. In case of an initial occupation of the right hand well, i.e. $N_{1R}\neq 0$ the situation becomes more complex. An example is shown in fig. \ref{fig:release}b where $N_{1R}=7\times 10^3$. Here $N_{1R}$ is not sufficiently large in order to provide confinement for all $m=0$ atoms, and a fraction of them escape to the $x>0$ halfspace. This emission recedes drastically as soon as a sufficient number of atoms has escaped and the number of $m=0$ atoms in the trapping region has become so small that confinement can eventually be granted by the $m=1$ atoms in the right well. This behavior is studied in more detail in fig. \ref{fig:release}c where we plot the time evolution of the number of escaped atoms $N_0^{esc}=\int_{x_{0}}^\infty|\Psi_{0}|^2$ for different values of the atom number $N_{1R}$ keeping $N_0=10^3$ and $N_{1L}=10^5$ constant. In the case at hand we define $x_{0}=80\, \mu\mathrm{m}$ thereby ensuring that the considered atoms are definitely not trapped any longer since the density of the $m=1$ component beyond $x_0$ is sufficiently small. The number of escaped atoms increases monotonously as time passes. For $t<60\, ms$ periods of a steep increase are followed by plateaus where almost no emission is observed. During the latter time intervals those atoms which are reflected from the $m=1$ wave packet localized in the left well destructively interfere with the atoms that initially went to the right, causing the emission to cease for a while. This is reflected in the pulsed release of $m=0$ atoms from the trapping region that is clearly seen in figs. \ref{fig:release}a and b. This situation is strongly reminiscent of a pulsed atom laser. For large times $N_0^{esc}/N_0$ saturates since the number of remaining atoms in the $m=0$ state is small enough to be confined. The number of confined atoms $N_{0}^{tr} = N_{0}-N_0^{esc}$ after $t=150\, ms$ is plotted in fig \ref{fig:release}d as a function of the occupation number of the right well for different $N_{0}$. As expected the final number of trapped atoms increases with increasing $N_{1R}$. The energy $E_{0}$ of the $m=0$ component is approximately proportional to the square of the occupation number $N_0$ in case of the nonlinear term being dominant. Furthermore, if one regards the $m=1$ component in the right well to act as a potential then the height $V_{1R}$ of this potential  is proportional to $N_{1R}$. The energy $E_{0}\propto N_0^2$ of trapped $m=0$ atoms should be less than the height of the potential $V_{1R}\propto N_{1R}$ created by the $m=1$ atoms. This yields the scaling $N_0^{tr}\propto \sqrt{N_{1R}}$ which is reproduced in fig. \ref{fig:release}d. The constant of proportionality depends on the initial occupation of the $m=0$ component. A larger initial occupation number leads to a larger amount of atoms with a small kinetic energy. As a consequence more atoms remain trapped reminiscent of the mechanism of evaporative cooling.

In the present study the focus was set on a 1d system but the results translate also to higher dimensions: For confinement in 2 and 3 dimensions ring \cite{Lesanovsky06,Hofferberth07,Lesanovsky06b} and shell-like \cite{Lesanovsky07} traps can be employed. Even more complex setups like arrays of BECs which provide a periodic 'soft' or 'hard' trapping potential are conceivable. Such a scenario, which is reminiscent of the self-assembled lattices presented in Ref. \cite{Pupillo08} can be realized via multi-well radio-frequency traps (see Ref. \cite{Lesanovsky07}).

\end{document}